\documentclass[aps, prl, superscriptaddress, preprintnumbers, twocolumn]{revtex4}

\usepackage{amsmath}	
\usepackage{amsthm}	
\usepackage{amssymb}	
\usepackage{datetime}
\usepackage{graphicx}
\usepackage{color}
\usepackage{verbatim}

\usepackage{hyperref}

\definecolor{grey}{rgb}{0.4,0.4,0.4}
\definecolor{dullmagenta}{rgb}{0.4,0,0.4}
\definecolor{darkblue}{rgb}{0,0,0.4}
\definecolor{midblue}{rgb}{0,0,0.5}
\definecolor{midred}{rgb}{0.5,0,0}
\definecolor{orange}{rgb}{1,0.5,0}
\definecolor{lightbrown}{rgb}{0.75,0.5,0.25}
\definecolor{tan}{cmyk}{0.14,0.42,0.56,0}
\definecolor{djunglegreen}{cmyk}{0.99,0,0.52,0}
\definecolor{lightgreen}{rgb}{0,1,0}
\definecolor{olivegreen}{cmyk}{0.64,0,0.95,0.40}
\definecolor{midgreen}{rgb}{0.0,0.675,0.0}
\definecolor{darkgreen}{rgb}{0,0.5,0}

\newcommand{\be}{\begin{equation}}
\newcommand{\ee}{\end{equation}}

\newcommand{\vs}{\vspace}

\renewcommand{\.}{\hspace{0.5mm}}

\newcommand{\Hrm}{\ensuremath{\mathrm{H}}}

\newcommand{\Vrm}{\ensuremath{\mathrm{V}}}

\newcommand{\crm}{\ensuremath{\mathrm{c}}}

\newcommand{\srm}{\ensuremath{\mathrm{s}}}

\newcommand{\Pcal}{\ensuremath{\mathcal{P}}}

\newcommand{\ie}{i.e.}

\newcommand{\cf}{cf.} 

\let\baraccent=\= 
\renewcommand{\=}[1]{\stackrel{#1}{=}} 

\theoremstyle{definition}

\theoremstyle{remark}

\settimeformat{ampmtime}

\usepackage{float}

\begin{document}

\title{On Ellipsoidal Collapse and Primordial Black-Hole Formation}

\author{Florian K{\"u}hnel}
\email{florian.kuhnel@fysik.su.se}
\affiliation{The Oskar Klein Centre for Cosmoparticle Physics,
	Department of Physics,
	Stockholm University,
	AlbaNova,
	SE--106\.91 Stockholm,
	Sweden}
	
\author{Marit Sandstad}
\email{marit.sandstad@astro.uio.no}
\affiliation{Nordita,
	KTH Royal Institute of Technology and Stockholm University,
	Roslagstullsbacken 23,
	SE--106\.91 Stockholm,
	Sweden}

\date{\formatdate{\day}{\month}{\year}, \currenttime}

\begin{abstract}
We reinvestigate gravitational ellipsoidal collapse with special focus on its impact on primordial black-hole formation. For a generic model we demonstrate that the abundance and energy density of the produced primordial black holes will be significantly decreased when the non-sphericity of the overdensities is taken into account.\\[-6mm]
\end{abstract}

\preprint{NORDITA-2016-14}

\maketitle

%
\noindent
The process of gravitational collapse is of utmost importance to our understanding of the Universe. From the formation of galaxies \cite{Blumenthal:1984bp,Kauffmann:1993gv}, clusters of galaxies \cite{White:1977jf,Davis:1985rj}, haloes \cite{Navarro:1995iw,Navarro:1996gj,Moore:1999nt,Bullock:1999he,Cooray:2002dia} (for a recent review see \cite{Frenk:2012ph}) or even to the possible formation of primordial black holes \cite{Carr:1974nx, Carr:2009jm, Khlopov:2008qy}, the nature of the collapse is crucial in determining characteristics like abundance, mass or shape.

In many cases, both for its calculational simplicity as well as being a reasonable first approximation, spherical symmetry has been an integral assumption to investigate gravitational collapse processes \cite{Misner:1964je, Penrose:1964wq, Penrose:1969pc}. Although in most cases initial non-sphericity is either small or eventually leads to (approximately) spherical objects, its effect might nevertheless be consequential. For instance, estimates of the abundance of small galactic haloes receive considerable corrections \cite{Sheth:1999su}, 
the formation of space-time singularities might be very different \cite{Shapiro:1991zza}, or, the precise geometric way in which collapse proceeds may lead to major intermediate deformations \cite{1965ApJ...142.1431L, Thorne:1972ji}.


The space of all possible shapes a collapsing overdensitiy might have is enormously large. Hence, one needs to focus on the most relevant structures. One of the simplest and most studied, deviating from spherical symmetry, is an ellipsoidal one, also because it gives a fairly good approximation to objects of many shapes. This has been and still is the focus of a vast amount of literature (\cf~\cite{1993A&A...280..351B, Eisenstein:1994ni, Bond:1993we, Monaco:1994ed, Monaco:1997cq, Sheth:1999su, Sheth:2001dp, Ohta:2004mx}), including the seminal work of Sheth, Mo and Tormen \cite{Sheth:1999su} who obtained a fitting formula for the mentioned collapse threshold which they found to be supported by numerical evidence. More recent evidence for the improvement of fits with an ellipsoidal collapse model can be found for instance in \cite{Robertson:2008jr, Angrick:2010qg}.

While most of the quoted references on ellipsoidal collapse deal with the formation of dark-matter haloes, the investigation of how the shape distribution of initial overdensities may affect the formation of primordial black holes is relatively modest. In \cite{Baumgarte:2015aza} the authors studied tri-axial collapse of black holes and critical collapse in a way which is relevant also for primordial black-hole formation, and in \cite{Clough:2016jmh} a non-spherical critical collapse was considered. The authors of \cite{Khlopov:1980mg, 1981SvA....25..406P, 1982SvA....26..391P} discussed effects of non-spherical geometry in the formation of primordial black holes if it occurred during an intermediate phase of matter domination due to a superheavy unified field theory particle in the very early universe. However, to the best of our knowledge, there has not been a thorough investigation of the effect of the abundance, or, the energy density of primordial back holes when lifting the spherical assumption on the overdensities in a general situation.

This is what we are going to study here. Specifically, we shall first argue that the ellipsoidal collapse threshold in the case of primordial black holes should be similar to that found in halo formation with only small deviation to the exact fitting constants. As we shall argue, the details of the radiation medium will essentially be contained in the spherical collapse threshold which has been obtained in the case for primordial black holes through detailed numerical studies \cite{Musco:2004ak, Musco:2008hv, Musco:2012au}. We will then investigate the influence of non-spherical effects on the final mass-density spectrum for a generic model of primordial black-hole formation.

To start, the ellipsoidal collapse threshold obtained by Sheth, Mo \& Tormen (\cf~Eq.~(3) of Ref.~\cite{Sheth:1999su}) for the case of halo collapse can be expressed via
\begin{align}
	\frac{ \delta_{\rm ec} }{ \delta_{\crm} }
		&\simeq
								1
								+
								\kappa\mspace{-1mu}
								\left(
									5\.e^{2}\.
									\frac{ \delta_{\rm ec} }{ \delta_{\crm}^{2} }
								\right)^{\!\!\gamma}
		=
								1
								+
								\kappa\mspace{-1mu}
								\left(
									\frac{ \sigma^{2} }{ \delta_{\crm}^{2} }
								\right)^{\!\!\gamma}
								,
								\label{eq:Sheth-Mo-Tormen}
\end{align}
with the threshold value for spherical collapse $\delta_{\crm}$, the ellipticity $e$, and the hight of the density power spectrum at the given scale $\sigma^{2}$. The parameter values for $\kappa$ and $\gamma$ were found to be $0.47$ and $0.62$, respectively. The final equality holds after inserting the most-probable (mp) value for the ellipticity, $e_{\rm mp} = \left(\sigma/\delta\right)/\sqrt{5}$. Actually, this is not entirely correct as the average value $\left< e \right> = 9 / \sqrt{10\pi\,}\.( \sigma / \delta ) \neq e_{\rm mp}$, as was pointed out in \cite{Angrick:2010qg}. There they found another set of values for $\kappa=0.6536$ and $\gamma=0.6387$ \cite{Angrick:2010qg}.

The above result \eqref{eq:Sheth-Mo-Tormen} has been derived and numerically confirmed for a very limited class of cosmologies only, mostly relevant to structure formation. This in particular does not include the case of ellipsoidal collapse in radiation domination, which is the most important one for primordial black-hole formation\cite{Carr:2009jm}. Below, for a Gaussian-distributed density-perturbation spectrum, we shall justify why the functional form of Eq.~\eqref{eq:Sheth-Mo-Tormen} is relevant also for the case of primordial black-hole formation. In fact, by giving an approximate physical argument in which both the derivation and the approximations made are dependent only upon the geometry of the collapse process, we suggest that the form Eq.~\eqref{eq:Sheth-Mo-Tormen} should indeed hold for ellipsoidal gravitational collapses in arbitrary environments.

%
In order to estimate the modification of the threshold in the case of non-spherical collapse, we note that in the ellipsiodal case, the collapse starts with the smallest axis first and after that the longer axes will collapse faster than linearly \cite{Bond:1993we}. It is hence suggestive from the mass dependence of the overdensity $\delta( M )$ that the density perturbation will be smaller by $\delta( \Delta M )$, where $\Delta M$ accounts for the difference in mass $M$ of a sphere to that of an ellipsoid.

If we, like in the halo-collapse models (investigated in Ref.~\cite{Sheth:1999su}), consider Gaussian-distributed overdensities, it can be shown that the expectation values for the shape of overdensities are given by \cite{Doroshkevich1970, Bardeen:1985tr, Bond:1993we},
\vs{-0.3mm}
\begin{align}
	\langle e \rangle
		&=
								\frac{3\.\sigma}{\sqrt{10\pi\,}\.\delta}
								\; ,
	\qquad
	\langle p \rangle
		=
								0
								\; ,
								\label{eq:<e><p>}
\end{align}
where the ellipticity $e$ and the prolateness $p$ are defined as 
\begin{subequations}
\begin{align}
	e
		&\equiv
								\frac{b^{2}
								\left(
									c^{2}
									-
									a^{2}
								\right)}
								{2
								\left(
									b^{2}\.c^{2}
									+
									a^{2}\.c^{2}
									+
									a^{2}\.b^{2}
								\right)}
								\; ,
								\\[2mm]
	p
		&\equiv
								\frac{b^{2}\.c^{2}
								-
								2\.a^{2}\.c^{2}
								+
								a^{2}\.b^{2}}
								{2
								\left(
									b^{2}\.c^{2}
									+
									a^{2}\.c^{2}
									+
									a^{2}\.b^{2}
								\right)}
								\; ,
\end{align} 
\end{subequations}
with $a$, $b$, and $c$ denoting the lengths of the three semi-major axes from the shortest to the longest \cite{Bardeen:1985tr}. The prolateness runs from $p = e$ in the maximally prolate case (\ie~one long axis and two equal-length short ones) to $p = -\.e$ in the maximally oblate case (\ie~one short axis and two equal-length long ones).

Since the collapse is initiated along the shortest axis, it may be compared to that of the largest sphere contained within it, \ie~one of radius $a$. Using the expectation value for the prolateness $\langle p \rangle = 0$ \footnote{In the original derivation of Eq.~\eqref{eq:Sheth-Mo-Tormen} (\cf~\cite{Sheth:1999su}) the term $e^{2}$ is really $e^{2} \pm p^{2}$ with $p^{2}$. The minus sign is used for the case of positive prolateness, the plus sign for negative prolateness. The formula \eqref{eq:Sheth-Mo-Tormen} is then obtained by assuming $p = 0$. However, we do not primarily expect a prolate spheroid with $p = e$ to collapse as a sphere. In addition the formula is not symmetric in $\pm\.p$, so even if $p = 0$ is the average value for $p$, inserting it does not give the average value for the critical threshold. Both of these facts can actually be seen by eye from the plot used to fit the original formula (\ie~Fig.~1 of Ref.~\cite{Sheth:1999su}). The error in these approximations, is however negligibly small in the small-$e$ regime.} and solving for the two longer axes in terms of the ellipticity $e$, we find that the volume of the ellipsoid, $V_{\!e}$, is
\begin{align}
	V_{\!e}
		&=
								V_{\!s}\;\frac{\left(1+3e\right)}{\sqrt{1-3e}}
								\; .
\end{align}

To get a rough functional form for the critical density threshold for the ellipsoid, we start by assuming a uniformly increased density within the ellipsoid and the sphere to which we compare it. This assumption is clearly false, but since the density distribution is assumed to be Gaussian, this mistake is independent of the collapsing medium and hence can be fitted equally well in the final version of the function. We also note that this effect will go in the direction of bringing the ellipsoidal density threshold closer to the spherical one than the estimate we obtain, which will hence serve as an upper bound.

The uniform-density assumption combined with the demand that the density threshold should be exceeded in the enclosed sphere, leads to an increase in mass $M_{e} = M_{s}\.V_{\!e} / V_{\!s}$. The density contrast associated with a given mass, roughly behaves as $\delta( M ) \sim M^{2 / 3}$ in the primordial black-hole case (\cf~\cite{Carr:1975qj}). This is also independent of the collapsing medium, however, might differ from the halo-collapse case and may lead to different numbers for the final fits. To first order in ellipticity, this leads to
\vs{-3mm}
\begin{align}
	\frac{ \delta_{\rm ec} }{ \delta_{\crm} }
		&\simeq
								\left(
									1
									+
									3\.e
								\right)
		=
								1
								+
								\frac{9}{\sqrt{10\.\pi\,}}\!
								\left(
									\frac{\sigma^{2}}{\delta_{\crm}^{2}}
								\right)^{\!\!1 / 2}
								.
								\label{eq:delta-ec-pre}
\end{align}
As we can see, this is roughly of the same form as the result of Sheth, Mo and Tormen (\cf~Eq.~\eqref{eq:Sheth-Mo-Tormen}). Although this estimate is based on assumptions which are only very approximate, we know that we can now refine this by substituting the numerical values of the multiplier $9 / \sqrt{10\pi\.}$ and exponent $1 / 2$ with constants $\kappa$ and $\gamma$ and then fit the function.

It is important to note that all the approximations, indeed all input to get this form, is medium independent. Though one input may be due to the special geometry of the primordial black-hole collapse from a horizon-size object, this will only give a possibility for slightly different fitting values for the primordial black-hole as compared to the halo case. However, we believe this difference is not tremendous and will mainly occur in the multiplier. Hence we believe that a fit made in the halo case will still be a good estimate for a halo collapse in any medium/Universe-model, and that the halo fit values will be a good first approximation also in the primordial black-hole case.

%
 
Above in our heuristic derivation of the elliptical threshold formula, we have also tried to convey that this shift is independent of the medium in which the collapse is taking place. Below we will go through derivations of the density thresholds for primordial black-hole formation which have been preformed under the assumption of spherical overdensities. We will argue that the geometry does not enter directly into these values in any other way than in that they represent values for $\delta_{\rm sc}$ which can then be fed into a relations like \eqref{eq:Sheth-Mo-Tormen} to find the more accurate threshold $\delta_{\rm ec}$ for a given average ellipticity.

In his pioneering work Carr \cite{Carr:1975qj} showed that, for an overdensity $\delta$ to collapse, its scale has to be roughly larger than the Jeans length, yielding
\begin{align}
	w
		&\lesssim
								\left(
									\frac{ M }{ M_{\Hrm} }
								\right)^{\!\!2 / 3}
								\delta_{0}
		\equiv
								\delta_{\Hrm}
								\; ,
								\label{eq:Carr-Threshold}
\end{align}
with $\delta_{0}$ being the original density contrast. Equation \eqref{eq:Carr-Threshold} also defines the density perturbation $\delta_{\Hrm}$ at horizon crossing. Above, $M_{\Hrm}$ is the initial mass inside the horizon, and $M$ is the mass contained in the initial volume of the overdensity. In the case of radiation domination Eq.~\eqref{eq:Carr-Threshold} yields the often-quoted value $\delta_{\crm} \simeq 1 / 3$.

This threshold has been found in numerical studies of gravitational collapse to be not quite accurate (\cf~\cite{Musco:2008hv}). In particular, in \cite{Harada:2013epa} a three-zone model calculation has been investigated within which the travel time of pressure waves that would cause expansion has been compared to the collapse time for the overdensity. From this the authors obtained a formula which fitted the results of the numerical studies \cite{Musco:2004ak, Musco:2008hv, Musco:2012au} more closely than the original threshold \eqref{eq:Carr-Threshold} obtained by Carr \cite{Carr:1975qj}.

For an ellipsoidal overdensity, the shortest axis of the overdensity will be first to cross the horizon and begin a possible collapse. When deriving the collapse threshold in the more proper way, as done by Harada and collaborators \cite{Harada:2013epa}, the shape of the overdensity does not matter, as it only compares the travel time of the pressure wave in the medium (along the shortest axis in the ellipsoidal case) to the collapse time along the same trajectory. Hence the dependence on the travelled distance is cancelled in the threshold which only depends on the equation-of-state parameter of the medium. The final form of the threshold reads \cite{Harada:2013epa}:
\begin{align}
	\delta_{\Hrm}
		&\geq
								\frac{ 3
									\left(
										1
										+
										w
									\right) }{
									5
									+
									3\.w }\.
								\sin^{2}\!
								\left(
									\frac{\pi\.\sqrt{w\,}}{ 
										1
										+
										3\.w }
								\right)
								\, .
								\label{eq:Harada-Threshold}
\end{align}
For a radiation medium this yields $0.41$ which is only $10$ -- $20\%$ off from the numerically well-established values of $0.45$ -- $0.47$ \cite{Musco:2008hv}.


%
In order to demonstrate the importance of considering ellipticity also in the case of primordial black holes, we will now apply the ellipsoidal threshold Eq.~\eqref{eq:Sheth-Mo-Tormen} in a specific, but quite generic model which has production of primordial black holes in a reasonable mass range. The precise nature of this model does not matter as {\it all} models will be affected in a similar way, irrespective of the mass range in which they are peaking.

We consider a running-mass model \cite{Stewart:1996ey
}. In the context of primordial black-hole production, these have been intensively studied (\cf~\cite{
Drees:2012sz, Leach:2000ea}). The perhaps simplest realisation may be expressed through the inflationary potential
\begin{align}
	\Vrm( \phi )
		&=
								\Vrm^{}_{\!0}
								+
								\frac{ 1 }{ 2 }\.m_{\phi}^{2}(\phi)\.\phi^{2}
								\, ,
								\label{eq:V(phi)}
\end{align}
with the constant $\Vrm_{\!0}$, and the scalar field $\phi$. There exists a plethora of embeddings of this model in various frameworks, such as hybrid inflation \cite{Linde:1993cn} for instance, which lead to different specific functions $m_{\phi}( \phi )$. These yield distinct expressions for the primordial density power spectra whose variance can be recast into the general form \cite{Drees:2011hb}
\begin{align}
	\big[ \sigma( k ) \big]^{2}
		&\simeq
								\frac{ 8 }{ 81 }\.\Pcal( k_{\star} )
								\bigg(
									\frac{ k }{ k_{\star} }
								\bigg)^{\!\! n( k ) - 1}
								\Gamma\!
								\left(
									\frac{ n_{\srm}( k ) + 3 }{ 2 }
								\right)
								\label{eq:sigma-running-mass}
								,
\end{align}
where the spectral indices $n( k )$ and $n_{\srm}( k )$ are given by
\begin{subequations}
\begin{align}
	n( k )
		&=
								n_{\srm}( k_{\star} )
								-
								\frac{ 1 }{ 2! }\.\lambda_{1}\.\ln\!
								\bigg(
									\frac{ k }{ k_{\star} }
								\bigg)
								+
								\frac{ 1 }{ 3! }\.\lambda_{2}\.\ln^{2}\!
								\bigg(
									\frac{ k }{ k_{\star} }
								\bigg)
								\notag
								\\[0.5mm]
		&\phantom{=\;}
								-
								\frac{ 1 }{ 4! }\.\lambda_{3}\.\ln^{3}\!
								\bigg(
									\frac{ k }{ k_{\star} }
								\bigg)
								+
								\ldots
								\; ,
								\label{eq:n-running-mass}
								\displaybreak[1]
								\\[1mm]
	n_{\srm}( k )
		&=
								n_{\srm}( k_{\star} )
								-
								\lambda_{1}\.\ln\!
								\bigg(
									\frac{ k }{ k_{\star} }
								\bigg)
								+
								\frac{1}{2}\.\lambda_{2}\.\ln^{2}\!
								\bigg(
									\frac{ k }{ k_{\star} }
								\bigg)
								\notag
								\displaybreak[1]
								\\[0.5mm]
		&\phantom{=\;}
								-
								\frac{1}{6}\.\lambda_{3}\.\ln^{3}\!
								\bigg(
									\frac{ k }{ k_{\star} }
								\bigg)
								+
								\ldots
								\; ,
								\label{eq:n-running-mass-2}
\end{align}
\end{subequations}
with real parameters $\lambda_{i}$, $i = 1, 2, 3$.

A convenient measure of how many primordial black holes are being produced can be given through the ratio of the energy density of primordial black holes (PBHs) by the total energy density,
\begin{align}
	\beta
		&\equiv
								\frac{ \rho_{\rm PBH} }
								{\rho_{\rm tot}}
								\; .
								\label{eq:beta-definition}
\end{align}
Employing the standard approximation of horizon-mass collapse \footnote{A more refined treatment of the collapse exhibits a so-called critical scaling spectrum for the primordial black-hole mass distribution \cite{Choptuik:1992jv, Koike:1995jm, Gundlach:1999cu, Gundlach:2002sx}, which has been numerically verified over many orders of magnitude in density contrast in \cite{Musco:2004ak, Musco:2008hv, Musco:2012au}. Recently, in \cite{Kuhnel:2015vtw}, we have investigated the effect of this critical scaling on the abundance of primordial black holes and found that it can in fact be very large. The recent work \cite{Clough:2016jmh} on crticality in elliptical collapse indicates that the critical collapse can be handled as has been done in \cite{Kuhnel:2015vtw}, only employing the ellipsoidal density threshold $\delta_{\rm ec}$ though we shall not do so here. Also, critical collapse and ellipticity are quite distinctive in nature. There are also many other subtleties which alter the production of primordial black holes such as non-Gaussianities \cite{Young:2015kda, Young:2015cyn}, Press-Schechter vs. peaks formalisms \cite{Young:2014ana}, separate universes \cite{Kopp:2010sh, Carr:2014pga} and the cloud-in-cloud problem \cite{Jedamzik:1994nr}.}, \ie~$M_{\rm PBH} \simeq M_{\rm H}$, and utilizing the Press--Schechter formalism \cite{1974ApJ...187..425P} leads to
\begin{align}
	\beta
		&\approx
								{\rm erfc}
								\bigg(
									\frac{ \delta_{\crm} }{ \sqrt{2\,}\.\sigma }
								\bigg)
								\; .
								\label{eq:beta}
\end{align}

\begin{figure}
	\centering
	\vs{-4.5mm}
	\includegraphics[scale=1, angle=0]{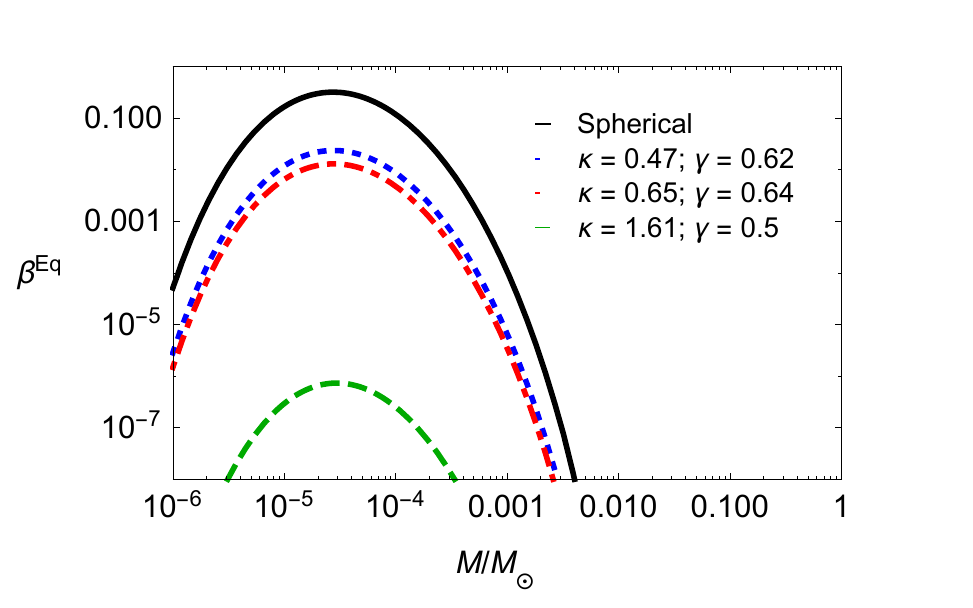}
	\caption{The relative energy density $\beta^{\rm Eq}$ for the running-mass case 
		at the time of radiation-matter equality and 
		as a function of $M / M_{\odot}$.
		The black, solid curve assumes spherical collapse 
		while the disconnected, colored ones represent non-zero ellipticity.
		See the figure legend and the main text for details.\vs{-3mm}}
	\label{fig:beta-Eq}
\end{figure}

In order to generate a significant fraction of primordial black holes in the mass range between $10^{-4}$ and $10^{-5}$ solar masses, we choose the parameter values $\lambda_{1} = 0.011$, $\lambda_{2} = 0.011$, $\lambda_{3} = - 0.0010975$, and display the subsequent results in Fig.~\ref{fig:beta-Eq}. Therein we show the ratio $\beta$ at the time of radiation-matter equality (superscript 'Eq') \footnote{One needs to time evolve $\beta$, being originally calculated at the time of formation, at each mass separately. In radiation domination, to a good approximation, $\beta$ grows linearly with the cosmic scale factor.} for three different cases: spherical collapse (black, solid), ellipsoidal collapse with the parameter values of Sheth, Mo and Tormen \cite{Sheth:1999su} (blue, dotted), and the ones obtained by Angrick and Bartelmann \cite{Angrick:2010qg} (red, dot-dashed). The values of the parameters $\kappa$ and $\gamma$ can be found in the legend of Fig.~\ref{fig:beta-Eq}. It can be observed that for the fitted parameters the suppression is approximately an order of magnitude. The corresponding curve representing the lower bound (\cf~Eq.~\eqref{eq:delta-ec-pre}) (green, dashed) lead to a suppression of several orders of magnitude below all the others, thereby also demonstrating the strong sensitivity of the ratio $\beta$ on the threshold $\delta_{\rm ec}$. For all graphs we use $\delta_{\crm} = 0.45$.

%
Summarized, our investigations show that even though non-spherical effects on the collapse threshold might seem to be small (generically being on the percent level), they can lead to tremendous suppression (of about an order of magnitude in most of the realistic cases) of the energy density of primordial black holes, as can be seen in Fig.~\ref{fig:beta-Eq}. As mentioned earlier, since in general neither the most-probable nor the average value for the ellipticity are zero, gravitational collapse to, in particular, black holes will {\it always} be influenced by non-sphericity.

\acknowledgments

We are indebted to Dominik Schwarz for drawing our attention to the importance of non-spherical effects, and also thank him, Thomas Baumgarte and Nico Wintergerst for useful discussions. F.K.~acknowledges supported from the Swedish Research Council (VR) through the Oskar Klein Centre.


\bibliography{refs}

\end{document}